# The effect of Ag addition on the superconducting properties of the FeSe system

E. Nazarova[1,2], K. Buchkov[1], S. Terzieva[1], K. Nenkov[2], A. Zahariev[1], D. Kovacheva[3], N. Balchev[1] and G. Fuchs[2]

[1]*Institute of Solid State Physics, Bulgarian Academy of Sciences, 1784 Sofia, Bulgaria*
[2]*Leibniz Institute for Solid State and Materials Research, (IFW Dresden), P.O. Box 2700116, D-01171 Dresden, Germany*
[3]*Institute of General and Inorganic Chemistry, Bulgarian Academy of Sciences, 1113 Sofia, Bulgaria*

**Abstract**

The tetragonal FeSe phase is an intensively investigated iron based superconductor. In this study we examined the influence of Ag addition on the superconducting properties of selenium deficient polycrystalline $FeSe_{0.94}$. The samples were obtained by solid state reaction and melting methods. XRD analysis shows the presence of tetragonal phase and EDX analysis establishes inhomogeneous Ag distribution in the grains. The superconducting properties were investigated by fundamental and third harmonic AC magnetic susceptibility. The intergranular critical current determined from AC magnetic susceptibility in the Ag doped sample is several times higher than that in the undoped one, obtained by melting at approximately the same temperatures. Intra-granular current is field independent up to almost 1000 Oe. Using the temperature dependence of third harmonic AC magnetic susceptibility at different DC magnetic fields, the irreversibility lines were obtained for all samples. It is found that Ag addition increased the irreversibility field in comparison with undoped melted and powder sintered samples. All results show that the Ag addition in selenium deficient ($FeSe_{0.94}$) samples leads to improvement of inter- and intra- granular properties: screening ability, pinning, activation energy and critical current and improves the irreversibility line.

## 1. Introduction

Since the discovery of superconductivity (SC) in the tetragonal FeSe phase [1] this compound is an intensively investigated iron based superconductor. This is because of its chemical and structural simplicity determined from the presence of only two elements. In spite of the fact that at ambient pressure its $T_c$ is about 8 K under the pressure $T_c$ increases up to 36.7 K [2, 3] thus showing that FeSe is a high temperature superconductor. This hints that there is a real potential for increasing of critical temperature and recently $T_c$ about 50 K was observed in single unit cell FeSe films on $SrTiO_3$ [4].

Compared to the cuprates, the Fe-based superconductors have some unique properties as low crystallographic anisotropy [5], high upper critical field [6], large Jc values weakly dependent on the field at low temperatures [7]. These characteristics prompt that the idea for wire fabrication is justified and its application in magnets and cables is expected. Soon after the discovery of oxy-pnictides the first LaFeAsOF wire has been fabricated by PIT technology [8]. Fe-based superconductors are hard and brittle materials and could not be plastically deformed. PIT method allows to start from powders packed in suitable metallic tube. This is relatively simple and low cost technics, but it is always based on polycrystalline materials. The main problem arising here is that weak intergrain connections limit the critical current density. In cuprates different metal additions have been used to improve the grain connectivity.

In this study we investigate the influence of Ag addition in FeSe on the superconducting properties and intergrain connections. The idea was conceived from our previous investigations concerning cuprates and recent publication [9], which claims that Ag in small concentration (about 5 %) enters into the crystal structure of $FeSe_{0.5}Te_{0.5}$, suppressing weakly $T_c$ and improving $J_c$ in magnetic field.

## 2. Experimental

The investigated samples with nominal composition $FeSe_{0.94}$ are obtained by two methods: solid state reaction and melting. Sample with 4 wt % Ag is also prepared by melting. The initial products Se, Fe and Ag powders with purity 99, 9% ; 99, 5% and 99,9% respectively are mixed and pressed into tablets in a glove box with Ar atmosphere. The tablets are put in silica tube, evacuated to $10^{-3}$ torr and sealed. The heat treatment is performed in vacuum furnace at $700^0$C for 8 hours. After a new grinding and pressing, the tablets are sealed in double evacuated quartz tubes for the melting process. The heat treatment is performed at $1050^0$C for 20 h. The synthesis is completed by an additional annealing at $400^0$C for 100 h followed by furnace cooling. For the solid state reaction method second synthesis is provided at $700^0$C for 24 h in single evacuated quartz tube and also annealed at $400^0$C for 36 h followed by furnace cooling.

Powder X-ray diffraction patterns are collected within the range from 5.3 to 80° 2θ with a constant step 0.02° 2θ on Bruker D8 Advance diffractometer with Cu Kα radiation and LynxEye detector. Phase identification is performed with the Diffrac*plus* EVA using ICDD-PDF2 Database.

The samples are characterized by dual beam scanning electron focused ion beam system (SEM/FIB LYRA I XMU, TESCAN), equipped with EDX detector (Quantax 200, Bruker)

The magnetic measurements are performed on the Quantum Design PPMS 9 T (at ISSP-Sofia), PPMS 14 T and MPMS (at IFW Dresden) in the temperature interval 2-300 K.

## 3. Results and discussion

The X-ray diffraction analysis of undoped samples shows the presence of tetragonal phase and a small peak, indicating Fe impurity. For melted sample the presence of hexagonal phase is observed. For sample with Ag addition all peaks are indexed as tetragonal phase and small amount of Ag and Fe impurity are also detected. SEM mapping of Ag doped sample (Fig.1) indicates inhomogeneous distribution of grains with different Ag content ranging from about 1 wt % up to 90 wt % as shown in the Table 1. In spite of the relatively small number of grains with exceptionally Ag content, most probably they are responsible for the small Ag peak in the diffractograms (not shown here).

*AC magnetic field amplitude dependence of magnetic susceptibility*

In granular superconductors double step $\chi_1(T)$ dependence is used to discriminate between the contribution of individual grains at high temperatures (intragranular superconductivity) and intergranular signal at low temperatures. It is impossible to realize such splitting from Fig.2, where the temperature dependences of the real and imaginary parts of fundamental AC magnetic susceptibility are presented for all investigated samples.

The critical temperature, $T_c$ for all obtained samples is ~ 8 K. The observed full screening below 4 K in Ag doped sample suggests the presence of good superconducting intergrain connections which are absent in both undoped samples. The bulk superconductivity was established in this sample by DC susceptibility measurements also (see inset in Fig.2). Further increasing of AC magnetic field amplitude (up to 15 Oe) broadens the superconducting transition more pronounced for the undoped sample and slightly decreases the $T_c$ . In Fig. 3 the effect of AC magnetic field amplitude on the imaginary part of $\chi_1(T)$ is demonstrated for two melted samples (Ag doped and undoped). The signal peaks at $T_p$ when the flux reached the center of the sample. The peaks broadened and shifted to lower temperatures when the AC magnetic field amplitude is increased. These changes are more dramatic for undoped sample where the full screening is not detected in $\chi'_1(T)$ dependence. For Ag doped sample, the intergarnular response is more stable and exists for higher AC magnetic field amplitudes. Approximating the samples as long cylinders with radius R and height h=10R, the intergranular critical current at $T_p$ was estimated from the relation $J_c(T_p)=H_{ac}/R$. As the sample dimensions are almost identical, the intergranular $J_c$ for Ag doped sample was found to be a few times higher than undoped specimen obtained by melting at approximately the same temperature ($J_c$(3.5K)=31.8 A/cm$^2$ for Ag doped sample and $J_c$(3.3K)=9.55 A/cm$^2$ for undoped one).

The full screening in Ag dopped samples supports the AC magnetic susceptibility harmonics registration. For real part of third harmonic a maximum is observed close to $T_c$ followed by large minimum at lower temperatures (Fig.4). Correspondingly the imaginary part shows small minimum near $T_c$ and large positive maximum at lower temperatures. In both third harmonics parts (real and imaginary) signals are influenced by magnetic field amplitude. The low temperature signals (the minimum of $\chi'_3$) reflecting the part of intergranular response are more influenced by magnetic field amplitude as a result of dissipative flux dynamical processes. In the undoped melted sample $FeSe_{0.94}$ the third harmonic signal as a function of temperature shows a similar curve shapes, but in smaller AC magnetic field amplitudes range (0.1 – 1.2 Oe). Based on these observations the following conclusion could be made: the intergranular critical current in Ag doped sample is higher and its flux pinning is improved.

*Frequency dependence of AC magnetic susceptibility*

The AC magnetic susceptibility response (fundamental and third harmonic) is influenced when the frequency is changed at fixed AC magnetic field amplitude (1 Oe). The peak intensity of $\chi_1''(T)$ dependencies rises, and this is accompanied by the shift to higher temperatures. The difference between the $T_p$ at the highest frequency (7777 Hz) and that at the lowest frequency (133 Hz) is $\Delta Tp = 0.87$ K. Similar behavior is characteristic for granular cuprates and was observed by us in Ca substituted YBCO polycrystalline samples [11]. However the observed displacement of the $\chi_1''(T)$ peak is smaller (0.45K-0.65K) at higher AC magnetic field amplitude (10 Oe) and frequencies up to 9999 Hz. By plotting $1/T_p$ vs. ln (f) for the Ag doped sample, a perfect linear dependence is obtained in the frequency range 133 – 5555 Hz. This behavior can be interpreted as Arrhenius plot $f=f_0\exp(-E_a/kT_p)$, where $E_a$ is the activation energy for thermally activated flux creep in the intergranular region and $f_0$ is a characteristic frequency. The obtained value for $E_a$ is 0.019 eV for $H_{ac}$=1 Oe and $f_0$ is of order of $10^8$ Hz as it is usually obtained. In undoped samples obtained by solid state reaction this tendency is also observed in spite of the fact that curves are not completed in the experimentally accessible temperature range. Following the procedure describe in [12] for $E_a$ determination we used the curves displacement with frequency on the level possibly very close to the maximum. The roughly estimated value in this case is an order of magnitude lower ($E_a$~0.0019 eV, at $H_{ac}$=1 Oe).

Careful examination of Cole-Cole plots of first harmonics shows that the signal increases with the frequency (Fig.5a). The observed peaks and followed valleys in real parts of third harmonic decrease when the frequency increases. For the imaginary part, the high temperature minimum is suppressed and the followed bump increases with the frequency. Cole-Cole plots of third harmonics are presented on Fig.5b. According to the model presented by [13] these characteristics of AC susceptibility suggest a vortex glass state in the sample. Closed loops tend to occupy the left half plane due to strong pinning and effective bulk screening [14]. Their area decreases with the frequency as a result of increasing linear processes. Nevertheless the strong frequency dependence observed in Ag doped $FeSe_{0.94}$ needs further investigation and clarification.

*DC magnetic field dependence of magnetic susceptibility*

For the Ag doped sample the positive signal of $\chi_1(T)$ in normal state is suppressed to zero and the full screening disappears by increasing the dc magnetic field ($H_{dc}$). The onset critical temperature at $H_{dc}$=13 T is about 4.5 K. It is worth mentioning that an effect of inversion is observed for $H_{dc}$=0 and $H_{dc}$=100 Oe. It was established that the superconducting transition at $H_{dc}$=100 Oe occurs at higher temperatures than that at $H_{dc}$=0. This could be a result from switching from intergranular to intragranular path but another explanation is not excluded also.

In order to compare both melted samples with and without Ag addition we performed similar investigations for $FeSe_{0.94}$ melted sample. Let's point out that $\chi_1''$ maximum at $H_{ac}$=1 Oe appeared at about 4 K and superposition of small $H_{dc}$ field (~500 Oe) shifts the maximum to the lower temperature out of the measuring scale. This comparison underlines the stability of intragranular signal in a wide temperature interval and magnetic field for Ag doped sample. It may be a result from better pinning due to Ag inclusions found to be present in many $FeSe_{0.94}$ grains.

In the Cole-Cole plots of Fig.6 the fundamental magnetic susceptibilities are presented for Ag doped sample at different $H_{dc}$ fields. The values of abscissas are normalized by using the maximum value of $|\chi_1'|$ for the given field, where the Meissner state is reached. $|\chi_1''|$ is also normalized to the corresponding maximum value.

It is seen that for $H_{dc}$-0-1000 Oe the $\chi_1'/\chi'_{1max}$ value is not zero at small fields ($H_{dc}$ < 1000 Oe), reflecting the signal from Fe impurities and vanish at higher fields ($H_{dc} \geq 1$ T). The steeper slope at $\chi_1' \rightarrow -1$ and maximum shift to the negative -1 side (when $H_{dc}$ increase) is an indication for flux creep existence in the sample [15]. For small fields (0-1000 Oe) the maximum occurs at $\chi_1'/\chi'_{1max} \approx 0.31$. Correcting this value by a small displacement (0.07) from zero because of the normal state positive $\chi_1'$ signal we obtained the exact value (0.38) predicted from Bean's critical state model [15]. Therefore, $J_c$ is almost field independent up to the $H_{dc}$=1000 Oe, while at higher fields the maximum is shifted to the higher values indicating the presence of flux creep in the grains.

The influence of dc magnetic field on the temperature dependence of third harmonic signal was examined too. Fig.7 shows the modulus of third harmonic signal as a function of temperature for Ag doped sample at different $H_{dc}$ fields. It is seen that for Ag doped sample the onset of the third harmonic signal remains at temperature higher than 2 K even at field of 13 T. The third harmonic signal is marking the reversibility-irreversibility boundary line and because of that its onset temperature at given field is used for irreversibility line (IL) determination. The experimental results are collected at minimal frequency 133 Hz in order to minimize influence of flux creep processes established from the analysis of Fig.6. In Fig.8 the irreversibility lines for both melted samples (Ag doped and undoped) are presented. The IL of undoped sample obtained by solid state reaction method is also given. Silver addition increases the irreversibility field: weakly compared with undoped melted $FeSe_{0.94}$ sample and more dramatic when compared with undoped sample produced by solid state reaction.

In conclusion, we demonstrated that samples grown from the melt show better superconducting properties than samples prepared by solid state reaction synthesis. Small addition of Ag (4%) has a positive effect on the inter- and intra- granular properties: screening ability, pinning, activation energy and critical current and slightly improves the irreversibility line.

**Acknowledgments**

This work was supported by the European Atomic Energy Agency (EURATOM) through the Contract of Association Euratom-INRNE.BG. One of the authors (E. N.) is grateful to the Leibniz Institute for Solid State and Materials Research, Dresden (especially to Prof. B. Holzapfel) for the invitation and possibility to perform the experiments.

Table 1. Results from Electron dispersive X-ray (EDX) analysis in 3 different points of Ag doped specimen

| Point | 1 | 2 | 3 | 1 | 2 | 3 |
|---|---|---|---|---|---|---|
| Element | Norm. wt % | | | Error, % | | |
| Ag | 1.50 | 7.16 | 91.80 | 1.0 | 0.4 | 2.7 |
| Fe | 58.28 | 41.83 | 2.66 | 1.3 | 1.3 | 0.1 |
| Se | 40.22 | 51.01 | 5.54 | 0.1 | 1.8 | 0.3 |

## Caption

Fig.1. SEM image of $FeSe_{0.94}$ +Ag melted sample (left panel) and mapping of Ag content in selected region (right panel).

Fig.2. Temperature dependence of real and imaginary parts of fundamental AC magnetic susceptibility for $FeSe_{0.94}$ samples obtained by solid state reaction (SSR) and melting and $FeSe_{0.94}$+Ag sample at 1 Oe AC magnetic field amplitude and frequency 133 Hz. Inset: temperature dependence of DC magnetic susceptibility in FC-ZFC regimes at H=100 Oe.

Fig.3. Temperature dependences of imaginary parts of fundamental AC magnetic susceptibility for different $H_{ac}$ magnetic field amplitudes for both melted samples. For the undoped sample the signals are increased 20 times.

Fig.4. Temperature dependence of real and imaginary pars of third harmonic AC magnetic susceptibility at different $H_{ac}$ magnetic field amplitudes for Ag doped sample.

Fig.5. Cole-Cole plots of fundamental (a) and third harmonic (b) AC magnetic susceptibility at constant AC magnetic field amplitude (1 Oe) and different frequencies.

Fig.6. Cole-Cole plots of normalized real and imaginary parts of fundamental magnetic susceptibility at different $H_{dc}$ fields and $H_{ac}$=1 Oe, f=133 Hz.

Fig.7. Modul of third harmonic signal as a function of temperature for Ag doped sample at different $H_{dc}$ field ranging from 0 to 13 T and small $H_{ac}$ amplitude of 1 Oe, f=133 Hz.

Fig.8 Irreversibility line determined from temperature dependence of modul of third harmonic AC magnetic susceptibility for Ag doped and undoped samples.

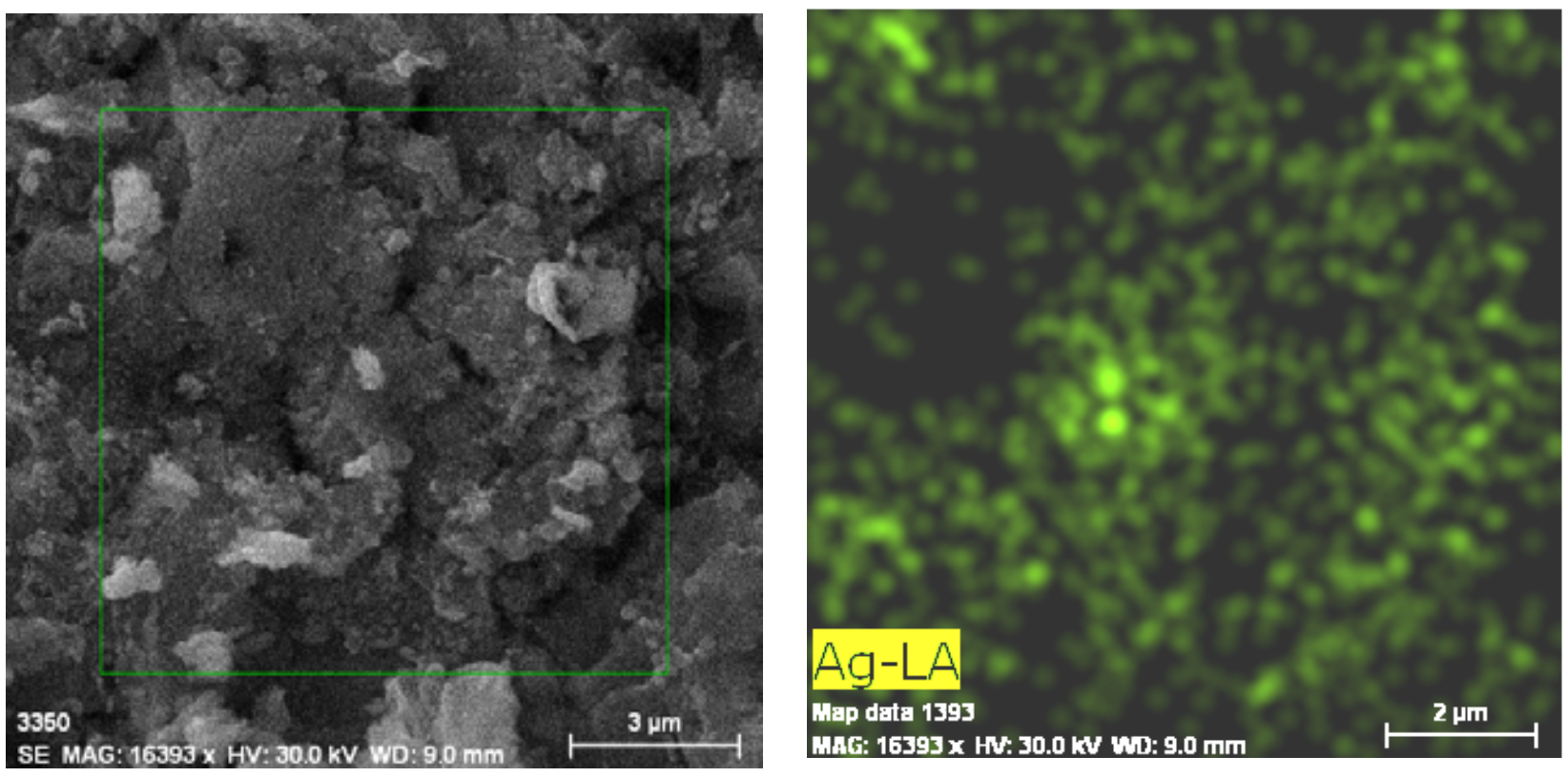

Figure 1.

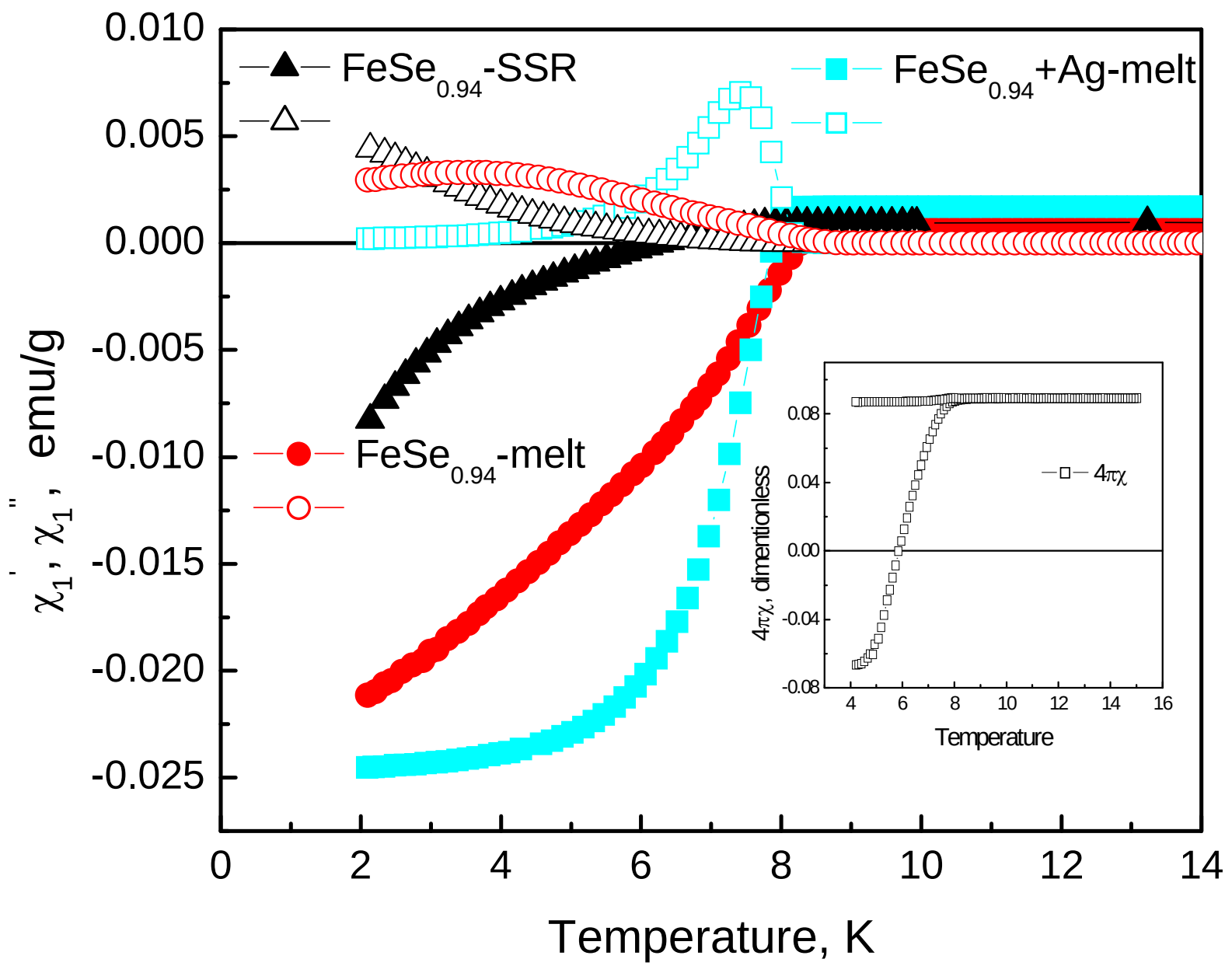

Figure 2.

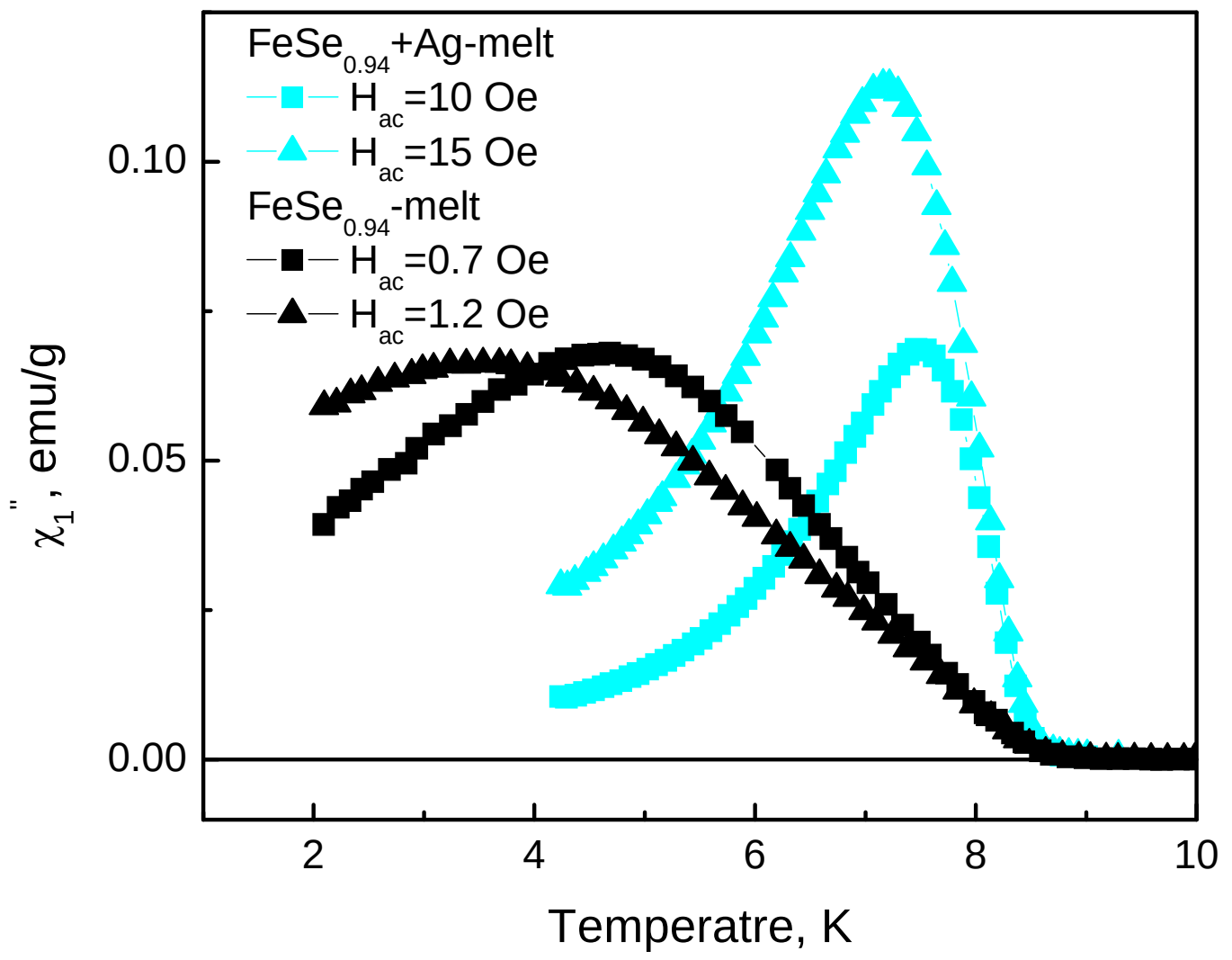

Figure 3.

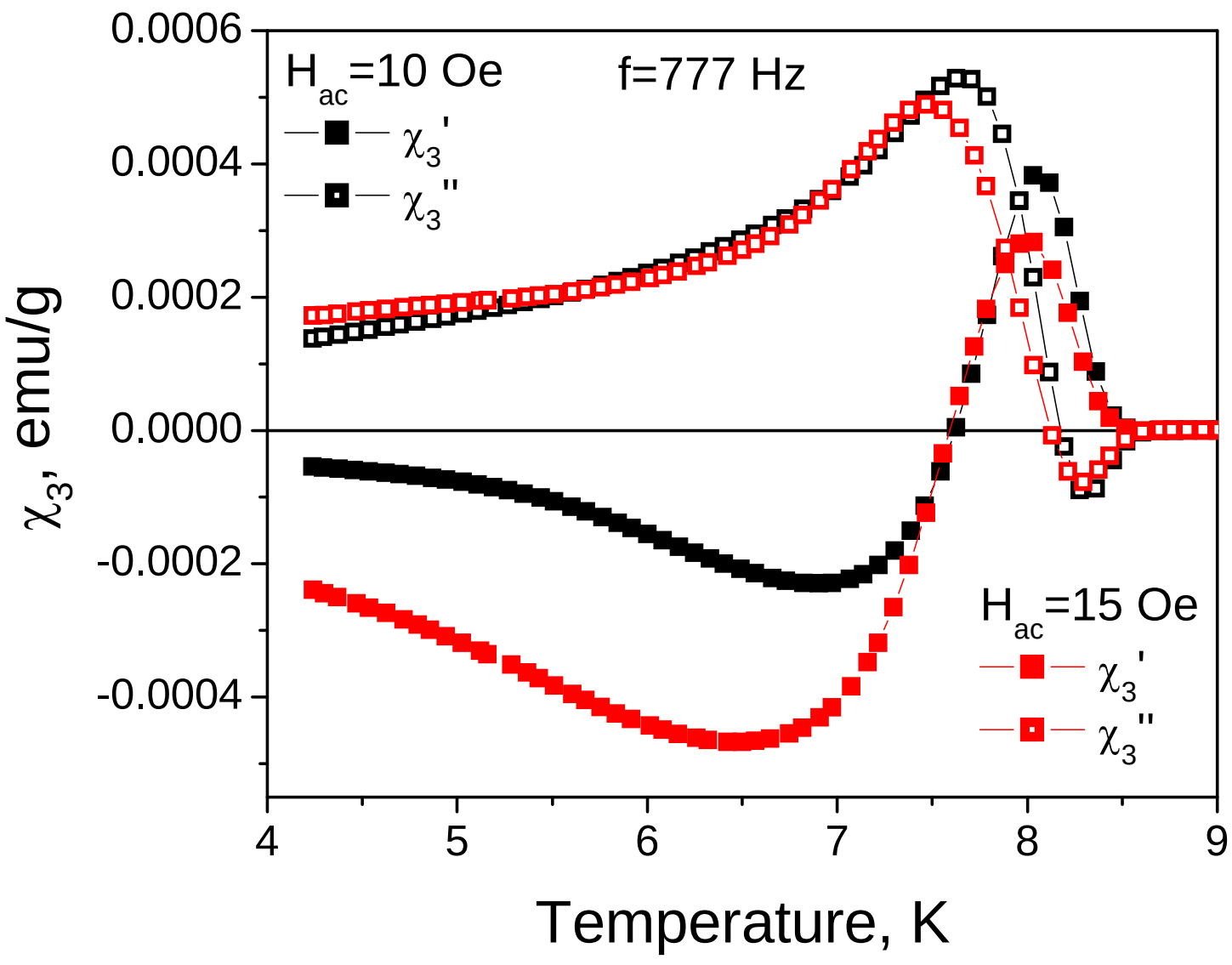

Figure 4.

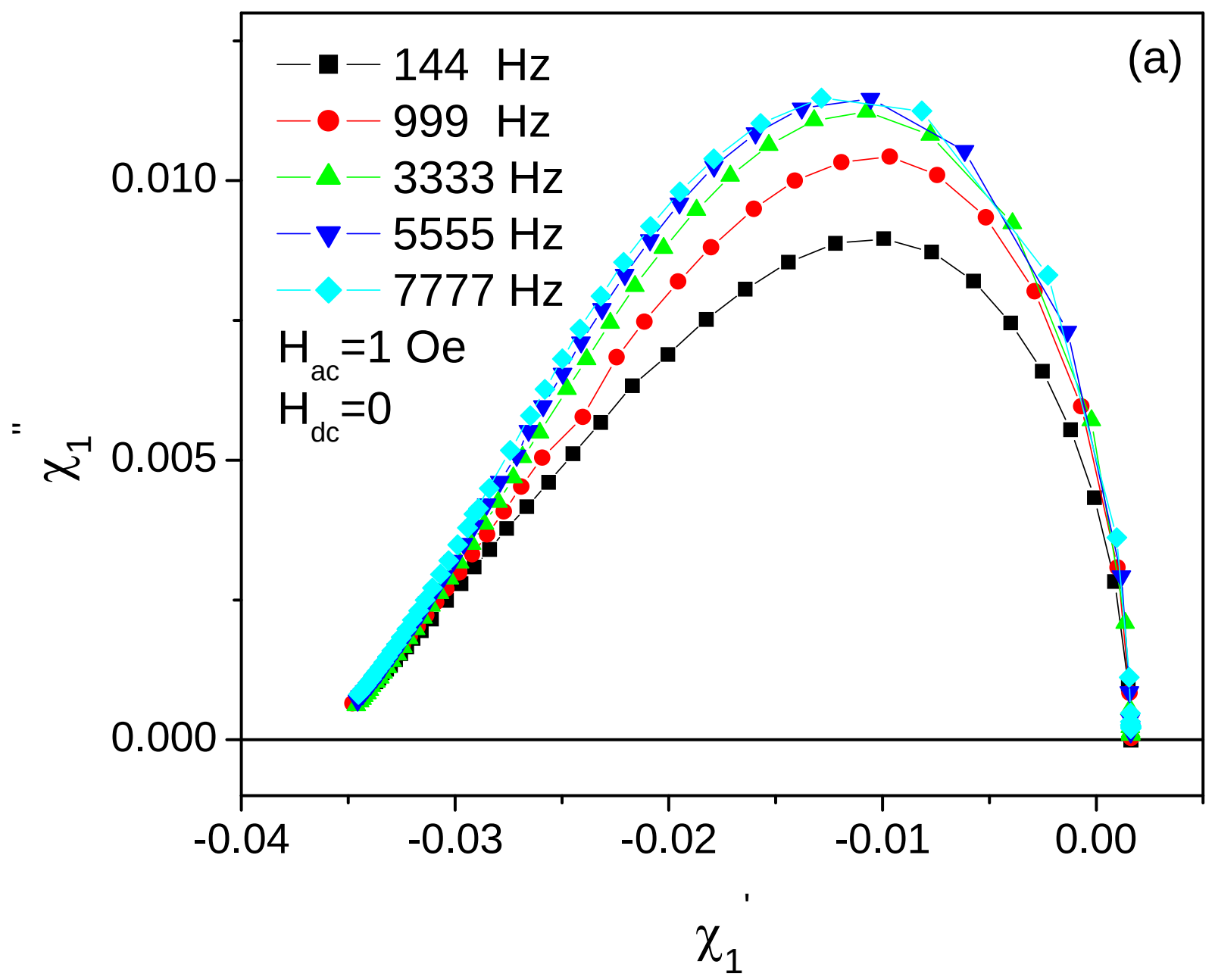

Figure 5a.

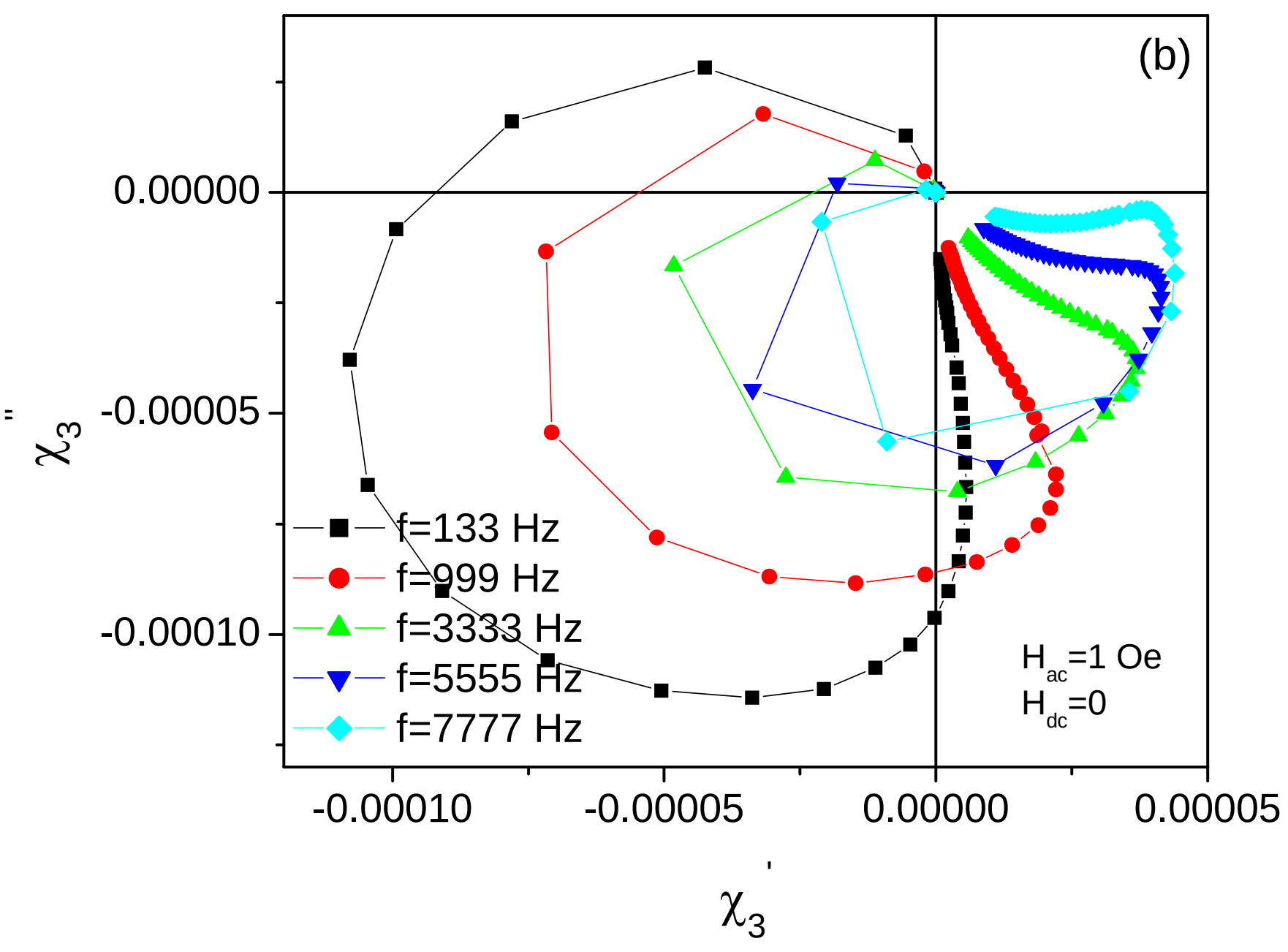

Figure 5b.

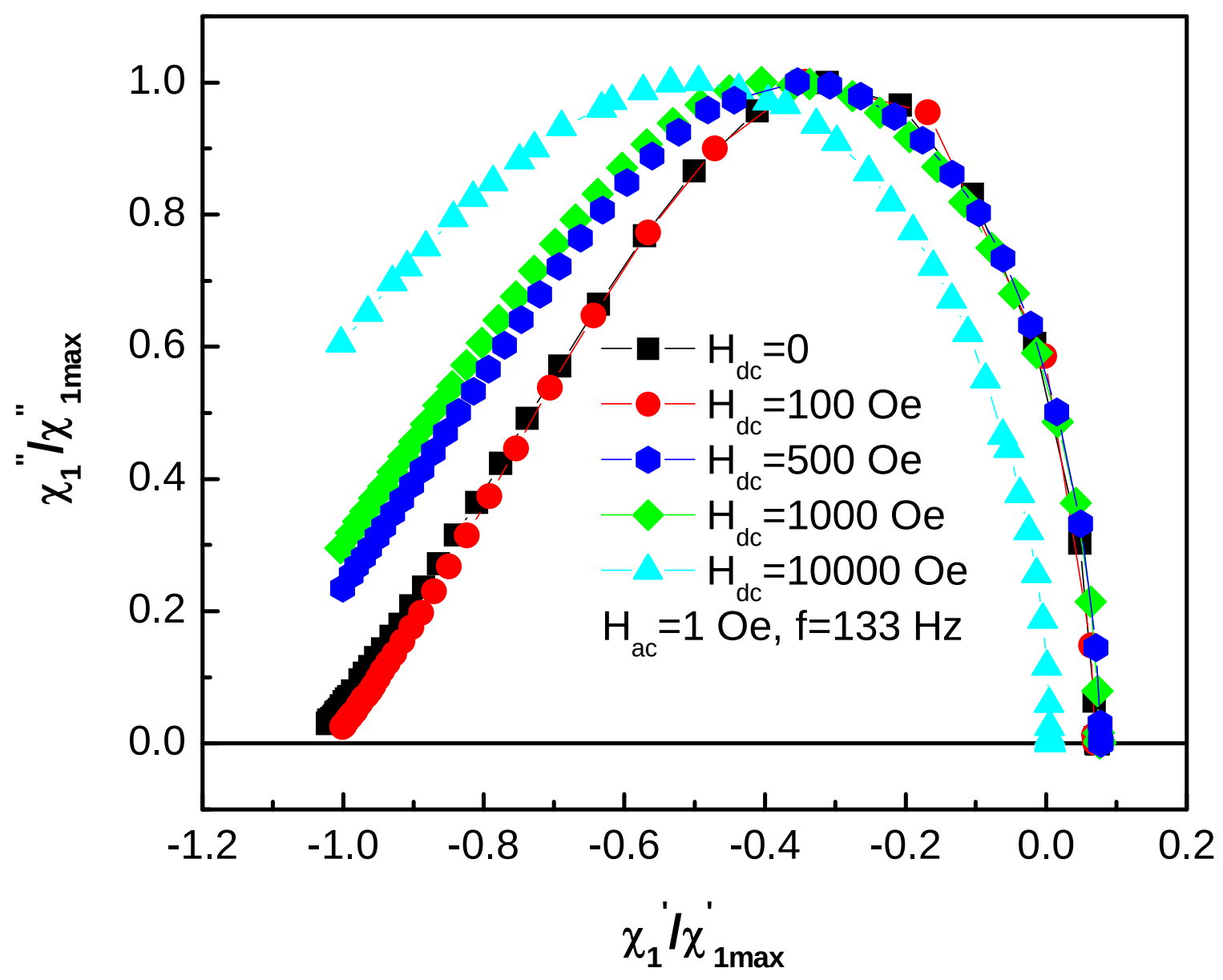

Figure 6.

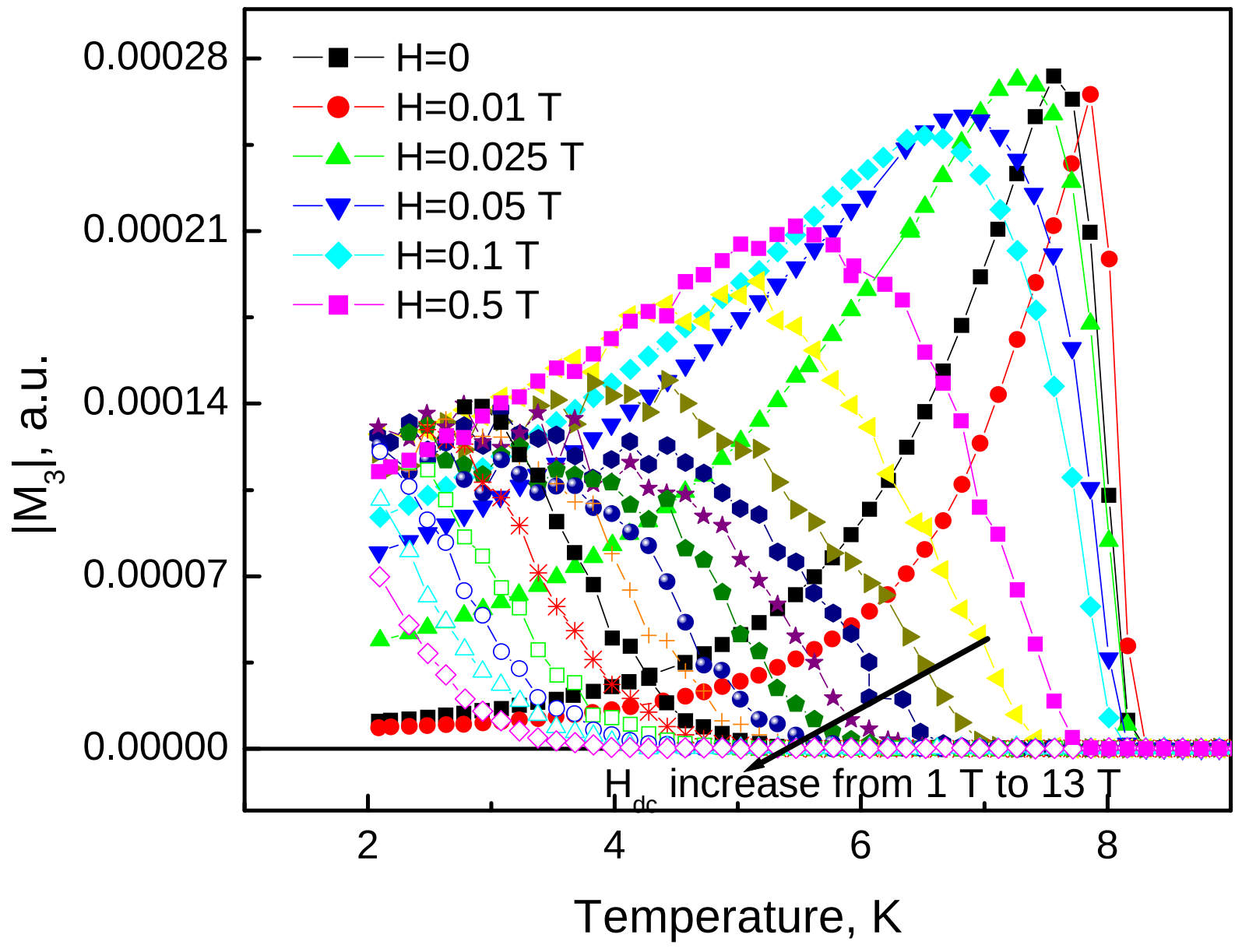

Figure 7.

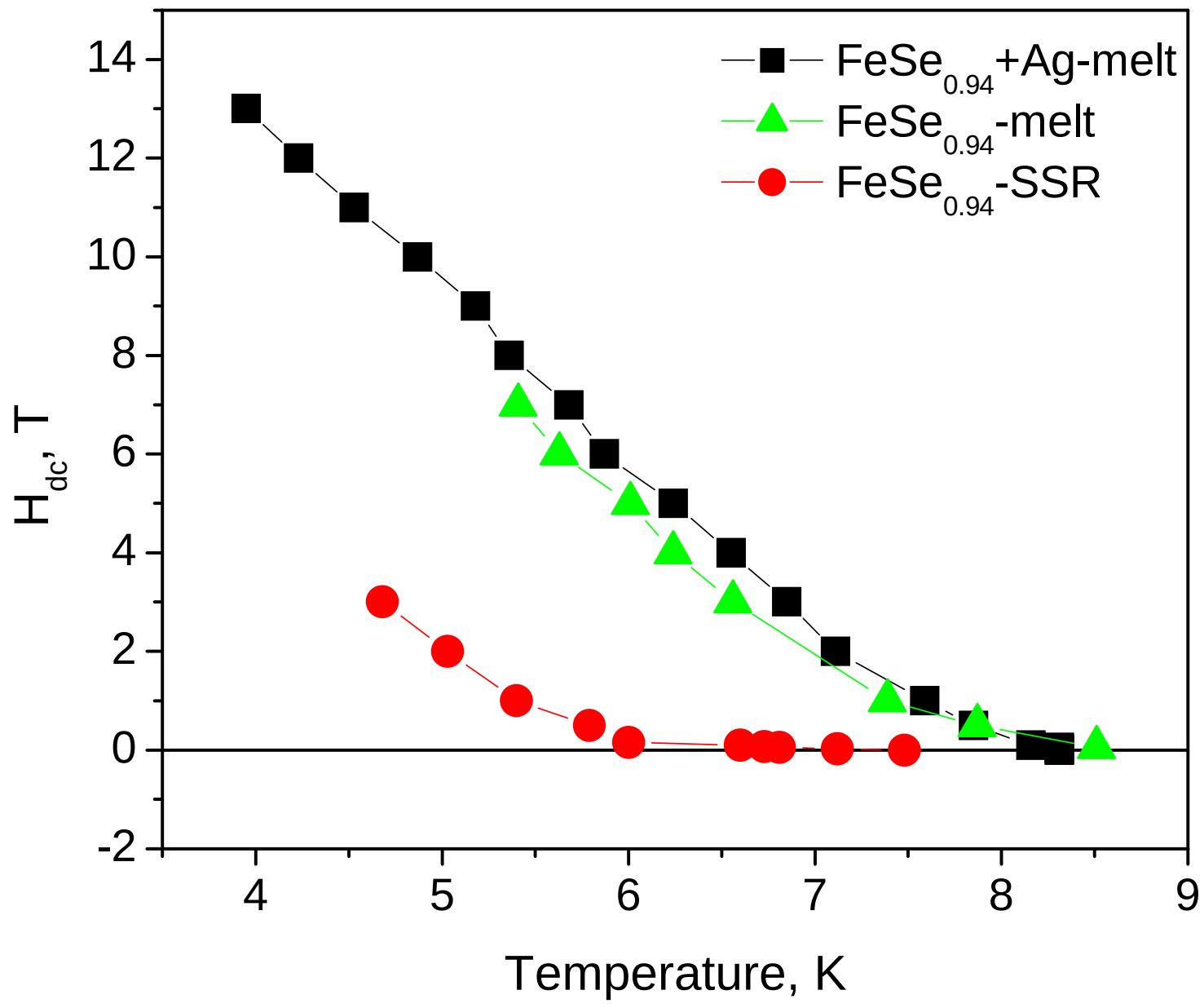

Figure 8.